# Pt-Bi Antibonding Interaction: The Key Factor for Superconductivity in Monoclinic BaPt$_2$Bi$_2$

Xin Gui,[†] Lingyi Xing,[§] Xiaoxiong Wang,[‡] Guang Bian,[&] Rongying Jin,[§] Weiwei Xie[†]*

[†]Department of Chemistry, Louisiana State University, Baton Rouge, LA 70803
[§]Department of Physics and Astronomy, Louisiana State University, Baton Rouge, LA 70803
[‡]Department of Applied Physics, Nanjing University of Science and Technology, Nanjing 210094
[&]Department of Physics and Astronomy, University of Missouri, Columbia, MO 65211

*Supporting Information*

**ABSTRACT:** In the search for superconductivity in BaAu$_2$Sb$_2$-type monoclinic structure, we have successfully synthesized a new compound BaPt$_2$Bi$_2$, which crystallizes in the space group $P2_1/m$ (S.G. 11; Pearson symbol $mP$10) according to a combination of powder and single crystal X-ray diffraction and scanning electron microscopy. Sharp electrical resistivity drop and large diamagnetic magnetization below 2.0 K indicates it owns superconducting ground state. This makes BaPt$_2$Bi$_2$ the first reported superconductor in monoclinic BaAu$_2$Sb$_2$-type structure, a previously unappreciated structure for superconductivity. First-principles calculations considering the spin-orbit coupling indicate that Pt-Bi antibonding interaction plays a critical role in inducing superconductivity.

The prediction of new superconductors is still an evasive goal for the solid state scientists. From the synthetic aspect, the most prominent challenge is to translate the physics of superconductivity into chemical requirements for new superconducting materials. A long-standing strategy for finding new superconductors is to assume that superconductivity exists in a favored structural motif. AM$_2$X$_2$ compounds (also referred as 122-type) with layered tetrahedral framework structures are widely accepted as the ideal charge-transfer-tunable systems.[1-2] In particular, materials in the body-centered tetragonal ThCr$_2$Si$_2$-type structure are well studied for their usual magnetic and superconducting properties.[3] A smaller number of superconducting materials crystallize in the primitive tetragonal CaBe$_2$Ge$_2$-type structure (S.G.129).[4-5] Although superconductivity was also discovered in the phase at the edge of phase transition between CaBe$_2$Ge$_2$- and BaAu$_2$Sb$_2$-type,[6] there is no evidence for superconductivity in monoclinic BaAu$_2$Sb$_2$–type structure.[7] From the structural viewpoint, the ThCr$_2$Si$_2$-type structure has two equivalent M$_2$X$_2$ layers per unit cell, while both the CaBe$_2$Ge$_2$- and BaAu$_2$Sb$_2$-type structures contain alternating M$_2$X$_2$ and X$_2$M$_2$ layers with a small electronegativity difference between M and X.[8] To search for superconductivity in BaAu$_2$Sb$_2$-type monoclinic structure, we choose an unexplored system based on Ba-Pt-Bi. For the first time, we have successfully synthesized BaPt$_2$Bi$_2$ compound. Our structural characterization reveals the BaAu$_2$Sb$_2$-type monoclinic structure. Both the electrical resistivity and magnetic susceptibility indicate superconducting transition at 2.0 K. We explore the origin of superconductivity by comparing its electronic structure with non-superconducting BaPd$_2$Bi$_2$.

Polycrystalline samples of Ba$_x$Pt$_2$Bi$_2$ ($x$= 1.0, 1.05, 1.1, 1.2, 1.3, 1.5) were prepared by the high-temperature solid-state synthetic method with elemental barium (>99%, rod, Alfa Aesar), powder platinum (99.98%, ~60 mesh, Alfa Aesar), bismuth (99.999%, lump, Alfa Aesar). Both platinum and bismuth are ground together and mixed with excessive barium pieces in the glovebox.[9] The mixture of three elements was pelletized and placed into an alumina crucible which was subsequently sealed into an evacuated (10$^{-5}$ torr) quartz tube. The following treatment was taken at 850 °C with a heating rate of 1 °C/min and then held for 2 days. After annealing, the quartz tube was quenched by air. The resulting sample was stored in the glovebox due to its sensitivity to both air and moisture. A Rigaku MiniFlex-600 powder X-ray diffractometer equipped with Cu K$_\alpha$ radiation ($\lambda$=1.5406 Å, Ge monochromator) was used to examine the phase information. The Bragg angle ranges from 0° to 90° in a step of 0.010° at the rate of 0.8°/min. The refinement was carried out by using LeBail mode with JANA 2006.[10-11] New ternary compound BaPt$_2$Bi$_2$ crystallizes in the monoclinic BaAu$_2$Sb$_2$-type structure, similar to BaPd$_2$Bi$_2$. A small amount of PtBi (<10%) coexists as the minor impurity phase according to the powder X-ray refinement.[12] For the X-ray powder diffraction patterns, as shown in Fig. 1(A), all scale factors and lattice parameters were refined using LeBail fitting, whereas the atomic sites or displacement parameters of all atoms were not refined. The resulting profile residuals $R_p$ is 9.06% with weighted profile residuals $R_{wp}$ 12.51%. The refined lattice parameters for BaPt$_2$Bi$_2$ (monoclinic symmetry, $a$ = 4.9694(2) Å, $b$ = 4.8474(2) Å, $c$ = 10.6463(4) Å, $\beta$= 91.912(2)°) showed a consistent result compared to the single crystal X-ray data. Analysis of samples, from both single crystals and SEM, confirms BaPt$_2$Bi$_2$ chemical composition, as shown in Table S1 and Figure S1. The high vacuum scanning electron microscope (SEM) (JSM-6610 LV) is used to determine the chemical composition. Samples were held on carbon tape before loading into the SEM chamber. Multiple points and areas were examined for each sample to get the Ba: Pt: Bi ratio.

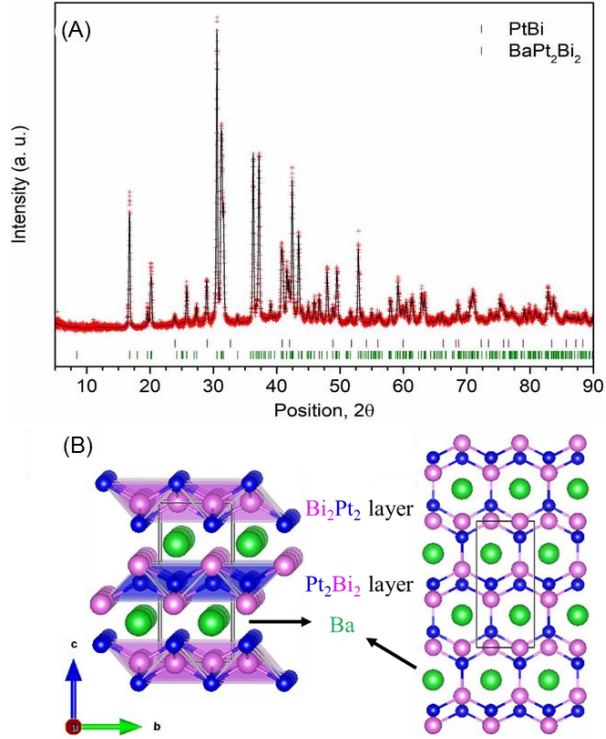

**Figure 1.** (A) Powder XRD pattern for $BaPt_2Bi_2$. The red points and black line represent observed and calculated intensities, respectively. Peak positions for $BaPt_2Bi_2$ and the PtBi impurity are labeled by green and purple vertical bars, respectively. (B)(Left) Crystal structure of $BaPt_2Bi_2$ emphasizing the $Pt_2Bi_2$ and $Bi_2Pt_2$ tetrahedral layers. (Right) View along the [100] direction with the focus on the Ba-filled Pt-Bi framework. (Green: Ba; Blue: Pt; Pink: Bi)

In order to avoid heterogeneity, more than twenty small crystals (~0.01×0.01×0.03 mm³) were chosen to take single crystal X-ray diffraction. The sample was analyzed by using a Bruker Apex II diffractometer with Mo radiation ($\lambda_{K\alpha}$= 0.71073 Å). Glycerol was used to protect the single crystals which were mounted on a Kapton loop. The measurement was taken in five different frames at 200 K to ensure accuracy. The direct methods and full-matrix least-squares on $F^2$ models with SHELXTL package were applied to solve the crystal structures.[13] To make data acquisition, Bruker SMART software was utilized, as well as making corrections for Lorentz and polarization effects.[14] Numerical absorption corrections were accomplished with *XPREP*, which is based on the face-index modeling.[15] Table 1 and Table 2 summarize the outcomes of structural investigation including the atomic positions, site occupancies, and isotropic thermal displacements.

**Table 1.** Single crystal refinement for $BaPt_2Bi_2$ at 200(2) K

| Refined Formula | $BaPt_2Bi_2$ |
| --- | --- |
| F.W. (g/mol) | 945.48 |
| Space group; Z | $P2_1/m$; 2 |
| a(Å) | 4.957 (2) |
| b(Å) | 4.822 (1) |
| c(Å) | 10.610 (3) |
| β (Å) | 92.00 (1) |
| V (Å³) | 253.5 (1) |
| Extinction Coefficient | 0.0029 (5) |
| θ range (deg) | 1.921 - 33.146 |
| No. reflections; $R_{int}$ | 2390; 0.0797 |
| No. independent reflections | 955 |
| No. parameters | 32 |
| $R_1$: $\omega R_2$ ($I>2\sigma(I)$) | 0.0691; 0.1442 |
| Goodness of fit | 0.997 |
| Diffraction peak and hole (e⁻/ Å³) | 6.440; -5.138 |

As shown in Fig. 1(B), $BaPt_2Bi_2$ adopts the $BaAu_2Sb_2$-type structure with the space group $P2_1/m$ (S.G. 11, Pearson Symbol *mP*10). The $BaAu_2Sb_2$-type structure has the lower symmetry, compared to the $CaBe_2Ge_2$ structure. From the crystal structure of $BaPt_2Bi_2$ in Fig. 1B (Left), $PtBi_4$ and $BiPt_4$ tetrahedral layers alternate along the *c* axis. The Pt-Bi distances in $BaPt_2Bi_2$ range from 2.74 to 2.83 Å, which are similar to the Pt-Bi distances in PtBi.[12] In spite of Pt-Bi interactions, half amount of Pt atoms form the zigzag chains is with the distance around 2.98 Å. The shortest Bi-Bi distance is 3.456(1) Å, which is not easily correlated with any bonding interaction. The electronegativity difference between Pt and Bi is too small to give rise to significant charge transfer between Pt and Bi. (Pauling Scale: 2.28 for Pt and 2.02 for Bi; Allen Scale: 1.72 for Pt and 2.01 for Bi) [16-17] Comparison to other compounds in $BaAu_2Sb_2$-type structure, as the atomic radii increase from Pd to Au on the M site and from Sb to Bi on the X site, the electronegativity difference between M and X keeps decreasing, the β angle keeps increasing as well (see Figure S1). Thus, the interplay of atomic size and electronegativity governs the distortion of the structure.

**Table 2.** Atomic coordinates and equivalent isotropic displacement parameters of $BaPt_2Bi_2$ system ($U_{eq}$ is defined as one-third of the trace of the orthogonalized $U_{ij}$ tensor (Å²)).

| Atom | Wyc. | Occ. | x | y | z | $U_{eq}$ |
| --- | --- | --- | --- | --- | --- | --- |
| Ba1 | 2e | 1 | 0.2375(5) | ¼ | 0.7437(3) | 0.0091(5) |
| Pt1 | 2e | 1 | 0.2555(4) | ¼ | 0.1248(2) | 0.0091(4) |
| Pt2 | 2e | 1 | 0.8227(4) | ¼ | 0.5023(2) | 0.0136(5) |
| Bi1 | 2e | 1 | 0.3035(3) | ¼ | 0.3830(2) | 0.0090(4) |
| Bi2 | 2e | 1 | 0.7491(3) | ¼ | 0.0021(2) | 0.0068(4) |

Fig. 2(A) shows the temperature dependence of magnetic susceptibility (χ) under different magnetic fields (H) between 1.9 and 2.5 K using a Quantum Design, Inc., superconducting quantum interference device (SQUID) magnetometer. The magnetic susceptibility is defined as χ = M/H where M is the measured magnetization in emu and H is the applied field in Oe. At H = 20 Oe, χ becomes negative with its magnitude increasing with decreasing temperature below $T_c$ ~ 2.0 K. This indicates that the system enters into the diamagnetic state below $T_c$. With increasing H, the diamagnetism is suppressed, suggesting that a superconducting transition exists at $T_c$. Indeed, the zero-field resistivity drop measured using a Quantum Design Physical Property Measurement System (PPMS) from 1.9 to 300 K, as shown in the inset of Fig. 2(B), confirms the superconducting transition. It also shows the linear behavior in above ~ 100 K, suggesting dominant electron-phonon scattering. At low temperatures, it slowly turns into $T^2$ dependence as demonstrated in the inset of Fig. 2(B), suggesting the strong electron-electron interaction.

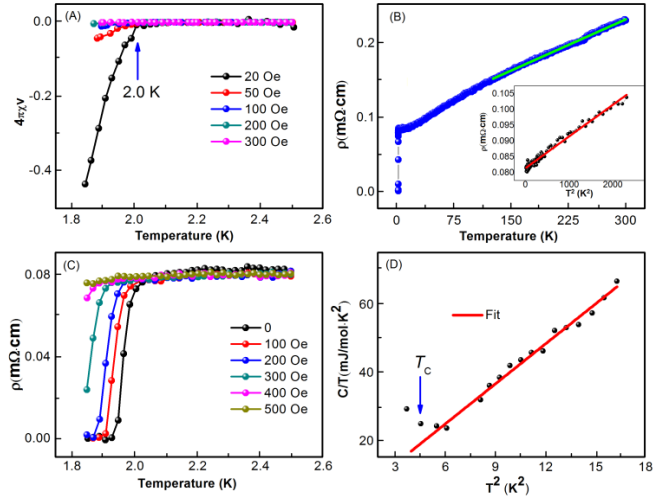

**Figure 2.** Physical measurements of the superconducting transition in BaPt$_2$Bi$_2$. (A) 4πχ$_v$ (T) measured under different applied magnetic field from 1.85 to 2.5 K with zero-field cooling and field cooling. (B) Resistivity vs. Temperature over the range of 1.8 to 300 K measured at zero applied magnetic field under 5000μA. Insert: Two different superconducting transition temperatures are determined as onset T$c$ and zero T$c$ as shown. (C) Resistivity *vs* Temperature under various fields. (D) Heat Capacity measurements from 2 to 4 K.

With the application of magnetic field, the transition is pushed towards lower temperatures and is more broaden in Figure 2(C). While magnetic susceptibility shows ~ 50% of Meissner volume fraction, low-temperature specific heat shows little anomaly at T$_c$. Fig. 2(D) exhibits temperature dependence of specific heat (C$_v$) plotted as C$_v$/T versus T$^2$ between 2.0 and 4.5 K. Note that C$_v$/T starts to depart from high-temperature linear dependence when approaching T$_c$, reflecting superconducting transition. Unfortunately, the complete characteristic specific heat peak is not observed due to an instability of temperature below 2.0 K, but the heat capacity starts jumping up from T$_c$= 2.3 K significantly. Nevertheless, by fitting the normal-state specific heat using C$_v$/T=γ+βT$^2$, we obtain the electronic specific heat coefficient γ ~ 1.06 mJ/mol·K$^2$ and phonon contribution β ~ 3.91 mJ/mol·K$^4$. The β value allows us to extract Debye temperature Θ$_D$ ~ 135 K. The value of the specific heat jump at T$_c$, estimated by the maximum value of specific heat observed at the low temperature limit of the measurement, is consistent with that expected from a weak coupling BCS superconductor. The ΔC$_{el}$/γT$_c$ per mole BaPt$_2$Bi$_2$ is larger than 1. This ratio is already within error of the BCS weak coupling value of 1.43 and is in the range observed for many superconductors. Thus, the observed superconductivity is intrinsic to BaPt$_2$Bi$_2$.

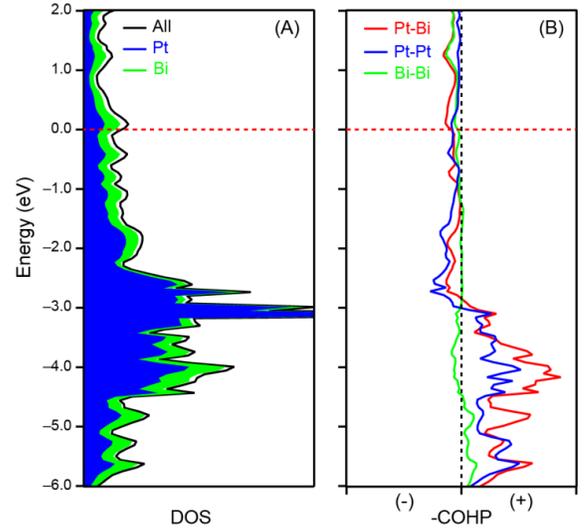

**Figure 3.** Partial DOS curves (A) and –COHP curves (B) of BaPt$_2$Bi$_2$ obtained from non-spin-polarization (LDA). (+ is bonding/ – is anti-bonding, E$_F$ is set to zero.)

To gain an intrinsic insight into the relationship between superconductivity and the electronic states of BaPt$_2$Bi$_2$, we investigate the electronic density of states (DOS) and Crystal Orbital Hamilton Population (–COHP) curves [18] for BaPt$_2$Bi$_2$ calculated by using Tight-Binding-Linear-Muffin-Tin-Orbital performed by using the Stuttgart code.[19] Figure 4(A) illustrates the total and partial density of states (DOS) for BaPt$_2$Bi$_2$. The DOS in the energy below –6.0 eV mainly consists of Bi *s*-orbital. The DOS in the energy range from –6.0 eV to +2.0 eV mainly consist of Bi *p*- and Pt *d*-orbital, in particular around Fermi level. Above +2.0 eV, the DOS originates from the Ba *d* and Pt *d* states. The significant DOS at E$_F$ is consistent with the metallic properties as we predicted from the structure and examined by experiments (electrical resistivity). A broad peak in the DOS in BaPt$_2$Bi$_2$ is often associated with a nearby structural, electronic, or magnetic instability such as superconductivity. Furthermore, the detailed atomic interactions in BaPt$_2$Bi$_2$ are analyzed through -COHP calculations in Fig. 3(B). According to the corresponding -COHP curves, the wavefunctions contributing to this peak around Fermi level have strong Pt-Bi and Pt-Pt antibonding characters. These features of the -COHP arise from structural influences on the orbital interactions in BaPt$_2$Bi$_2$. Thus, according to the DOS and -COHP curves, BaPt$_2$Bi$_2$ is susceptible toward either a possible structural distortion by disrupting the antibonding Pt-Bi and Pt-Pt orbital interactions at the Fermi level or toward superconductivity.

The other question, why other monoclinic 122-type compounds, such as BaPd$_2$Bi$_2$, do not show superconductivity, is still a puzzle. To unravel it, we calculated the band structure without/with spin-orbit coupling of BaPt$_2$Bi$_2$ using the VASP code [20] with generalized gradient approximation (GGA) schemes.[21] Structural lattice parameters obtained from experiments are used for our calculations. The spin-orbit coupling is included on all atoms. The projector augmented wave method is applied, and the energy cutoff is 400 eV. [22-23] Reciprocal space integrations are completed over an 11×11×5 Monkhorst-Pack *k*-points mesh.[24] The broad peak in DOS is due to the presence of saddle points in the electronic structure at the Z and X points in the Brillouin zone. These saddle points near $E_F$ are often proposed to be important for yielding superconductivity in a variety of materials. By comparing the Fermi surfaces of non-superconducting $BaPd_2Bi_2$ and superconducting $BaPt_2Bi_2$ in Figure 4(A) and (B), it is clearly shown that the hybridization between Bi-*p* and Pt-*d* orbit leads to the saddle point at Z-point and high DOS at the Fermi level. On the other hand, $BaPd_2Bi_2$ in the same monoclinic $BaAu_2Sb_2$ structure doesn't possess such Pd-Bi interactions, and thus the DOS is low compared to that of $BaPt_2Bi_2$. The "rigid band" model proposes that the structure type and electron count are paramount factors in superconductivity while the actual atomic configuration is generally not a primary consideration.[25] But, from the comparison of non-superconducting $BaPd_2Bi_2$ and superconducting $BaPt_2Bi_2$, the "critical pairs" of atoms, which gives the exactly right charge transfer between the atoms, are a crucial factor for inducing superconductivity.

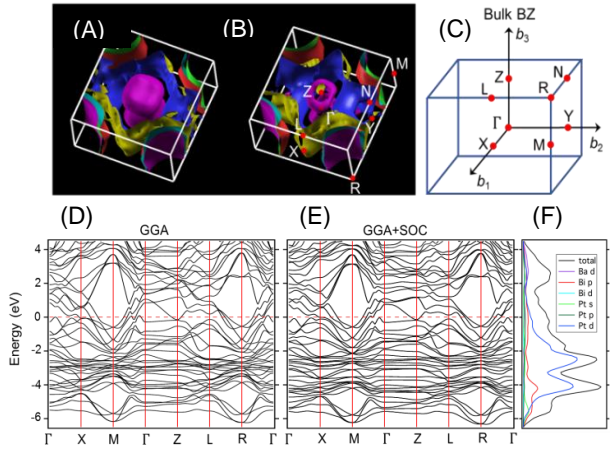

**Figure 4.** Fermi surfaces (FSs) for (A) non-superconducting $BaPd_2Bi_2$ and (B) superconducting $BaPt_2Bi_2$. (C) Brillouin zone of the $BaPt_2Bi_2$ lattice. Results of the electronic structure calculations for $BaPt_2Bi_2$ without (D)/with (E) spin-orbit coupling. (F) Total and partial DOS curves and band structure curves obtained from GGA calculations with the inclusion of SOC.

Herein, we reported the first ternary phase in Ba-Pt-Bi system, monoclinic $BaPt_2Bi_2$, which has been successfully synthesized. The crystal structure has been characterized by XRD measurements and the electronic structure has been obtained by first-principles methods. A combination of resistivity, magnetic susceptibility, and heat capacity measurements unambiguously shows that $BaPt_2Bi_2$ is a superconductor with $T_c \sim 2.0$ K. The electronic structure analysis of non-superconducting $BaPd_2Bi_2$ and superconducting $BaPt_2Bi_2$ illustrates that the Pt-Bi pair is a good pair for superconductivity in intermetallic compounds. The work presented here indicates the critical charge-transfer pairs can be a useful design paradigm for new superconductors.

**Acknowledgement**

X.G. and W.X. were supported by the Board of Regents Research Competitiveness Subprogram (RCS) under Contract Number LEQSF(2017-20)-RD-A-08 and LSU-startup funding. L.X. and R. J. were supported by the DOE DE-SC0016315. X. W. was supported by the National Natural Science Foundation of China (No. 11204133), the Fundamental Research Funds for the Central Universities (No.30917011338). G. B. was supported by University of Missouri startup funding.

**Reference**
[1]. Hosono, H.; Tanabe, K.; Takayama-Muromachi, E.; Kageyama, H.; Yamanaka, S.; Kumakura, H.; Nohara, H.; Hiramatsu, H.; Fujitsu, S. *Sci. Technol. Adv. Mater.* **2015**, *16*, 033503.
[2]. Kamihara, Y.; Watanabe, T.; Hirano, M.; Hosono, H. *J. Am. Chem. Soc.* **2008**, *130*(11), 3296–3297.
[3]. Rotter, M.; Tegel, M.; Johrendt, D. *Phys. Rev. Lett.* **2008**, *101*(10), 107006.
[4]. Kudo, K.; Nishikubo, Y.; Nohara, M. *J. Phys. Soc. Jpn.* **2010**, *79*, 123710.
[5]. Shelton, R. N.; Braun, H. F.; Musick, E. *Solid State Comm.* **1984**, *52*(9), 797-799.
[6]. Xie, W.; Seibel, E. M.; Cava, R. J. *Inorg. Chem.* **2016**, *55*(7), 3203-3205.
[7]. Frik, L.; Johrendt, D.; Mewis, A. *Z. Anorg. Allg. Chem.* **2006**, *632*, 1514-1517.
[8]. Zheng, C.; Hoffmann, R.; Nesper, R.; Schnering, H.-G. V. *J. Am. Chem. Soc.* **1986**, *108*(8), 1876–1884.
[9]. Jia, S.; Jiramongkolchai, P.; Suchomel, M. R.; Toby, B. H.; Checkelsky, J. G.; Ong, N. P.; Cava, R. J. *Nat. Phys.* **2011**, *7*, 207-210.
[10]. Bail, A. L.; Duroy, H.; Fourquet, J. L. *Mat. Res. Bull.* **1988**, *23*, 447-452.
[11]. Petricek, V.; Dusek, M.; Palatinus, L. *Z. Kristallogr.* **2014**, *229*(5), 345-352.
[12]. King-Smith, R. D.; Vanderbilt, D. *Phys. Rev. B* **2003**, *47*(3), 1651.
[13]. Sheldrick, G.M. *Acta Crystallogr. C* **2015**, *71*(1), 3-8.
[14]. *Smart*. Bruker AXS Inc.: Madison, WI, USA, **2012**.
[15]. Walker, N.T.; Stuart, D. *Acta Cryst.* **1983**, A39, 158-166.
[16]. Pauling, L. *J. Am. Chem. Soc.* **1932**, *54*(9), 3570–3582.
[17]. Allred, A. L. *J. Inorg. Nucl. Chem.* **1961**, *17*, 215–221.
[18]. Dronskowski, R.; Bloechl, P. E. *J. Phys. Chem.* **1993**, *97*(33), 8617-8624.
[19]. Krier, G.; Jepsen, O.; Burkhardt, A.; Andersen, O. K. *TBLMTO-ASA Program*, 4.7 ed., Stuttgart, Germany, **1995**.
[20]. Tao, J.; Perdew, J.P.; Staroverov, V.N.; Scuseria, G.E. *Phys. Rev. Lett.* **2003**, *91*(14), 146401.
[21]. Kresse, G.; Furthmüller, J. *Comput. Mater. Sci.* **1996**, *6*, 15-50.
[22]. Perdew, J. P.; Burke, K.; Ernzerhof, M. *Phys. Rev. Lett.* **1996**, *77*, 3865-3868.
[23]. Blöchl, P. E. *Phys. Rev. B* **1994**, 50, 17953-17979.
[24]. Kresse, G.; Joubert, D. *Phys. Rev. B* **1999** 59, 1758-1775.
[25]. Matthias, B. T. *Phys. Rev.* **1953**, *90*, 487.

TOC Graphic Only

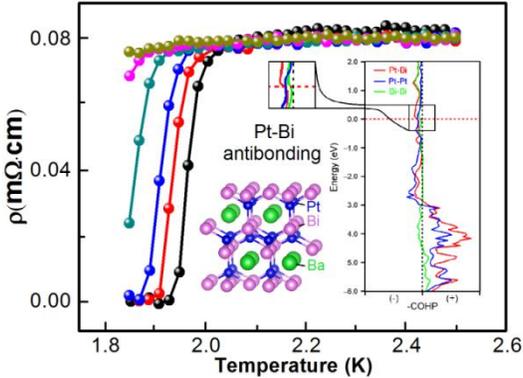